\documentclass[conference]{IEEEtran}
\usepackage{cite}      
\usepackage{graphicx}  
\usepackage{subfigure}

\begin{document}

\title{Connecting Spiking Neurons to a Spiking Memristor Network Changes the Memristor Dynamics.}

\author{
	\authorblockN
	{
	Deborah Gater\authorrefmark{3},
	Attya Iqbal\authorrefmark{2}\authorrefmark{4},
	Jeffrey Davey\authorrefmark{2}\authorrefmark{4} and
	Ella Gale\authorrefmark{1}\authorrefmark{2}\authorrefmark{5}\authorrefmark{4}
	}
	\authorblockA{
			\authorrefmark{3}Khalifa University of Science,	Technology and Research, 
                        Al Saada Street, P.O. Box 127788, 
			Abu Dhabi,	UAE\\
					}
	\authorblockA{\authorrefmark{4}Centre for Research in Biosciences}
			\authorblockA{
			\authorrefmark{5} Centre for Unconventional 	Computing and Bristol Robotics Laboratory}	
	\authorblockA{
			\authorrefmark{2} University of the West of 								England, Coldharbour Lane, Bristol, UK, BS16 1QY
			}
		\authorblockA{
			\authorrefmark{1}Email: ella.gale@uwe.ac.uk}	
}

\maketitle

\begin{abstract}
Memristors have been suggested as neuromorphic computing elements. Spike-time dependent plasticity and the Hodgkin-Huxley model of the neuron have both been modelled effectively by memristor theory. The d.c. response of the memristor is a current spike. Based on these three facts we suggest that memristors are well-placed to interface directly with neurons. In this paper we show that connecting a spiking memristor network to spiking neuronal cells causes a change in the memristor network dynamics by: removing the memristor spikes, which we show is due to the effects of connection to aqueous medium; causing a change in current decay rate consistent with a change in memristor state; presenting more-linear $I-t$ dynamics; and increasing the memristor spiking rate, as a consequence of interaction with the spiking neurons. This demonstrates that neurons are capable of communicating directly with memristors, without the need for computer translation.
\end{abstract}

\IEEEpeerreviewmaketitle

\section{Introduction}
The human body is partly electrical in nature, with nerve impulses being electrical signals, cardiac motion being controlled by electrical impulses and the brain `computing' via electrical spikes (both current spikes across synapses and voltage spikes along axons). The development of systems that are capable of reliable electrical interface with the brain and that can be incorporated into portable, wearable devices, could allow the creation of cognitive prosthetics to combat a number of serious neurological conditions resulting from disease, aging or injury.

The memristor~\cite{14} is a resistor that possesses a memory. Synapse action~\cite{71,216,236} is memristive and the ion channels involved in neuronal signal transport are biological memristors~\cite{247,248}. Most research has focussed on using memristors as computer memory and investigating the distinctive pinched hysteresis curve seen in response to sinusoidal voltage waveforms~\cite{15}. However, the d.c. response of the memristor is a current spike that contains the short-term memory of the memristor~\cite{Spcj}; on a single memristor these spikes can be used to compute in a Boolean fashion~\cite{P0c}, and networks of memristors exhibit brainwave-like dynamics and bursting spikes~\cite{c0c}. The intriguing similarity between memristor network dynamics, and brainwaves and neuronal spiking pattens suggests that the memristor could be a good choice for either copying the brain (i.e. creating a biomimetic neuromorphic computer) or interfacing with it. In this paper we present preliminary results from 
an ongoing investigation if and how memristor spikes can affect neuronal cell spikes (i.e. can memristors act as an input to a living neuronal network) and if the spiking cells can alter the memristor dynamics -- the latter question concerns us here. Various studies~\cite{Spcj,P0c,c0c} have shown that the memristor spikes are reproducible and can interact through time on a single memristor and/or through space in a network of memristors. Here we show that neuronal cell spiking can affect the spiking properties of an attached memristor network. 

The memristors used in this study~\cite{260} are flexible (which may make them more suitable for interfacing with cells in future applications), operate with physiological currents~\cite{243}, and are made of spun-on titanium dioxide sol-gel~\cite{M0} between sputtered aluminium electrodes. We postulated that the current spikes could trigger spiking in cells, but it was not known if neuronal cell voltage spikes would affect the memristors: usually the lowest voltage the memristors are tested with is 0.01V and $I-V$ curves typically range over $\pm1V$.

\section{Methodology}

\begin{figure}
\centering
\includegraphics[width=3.5in]{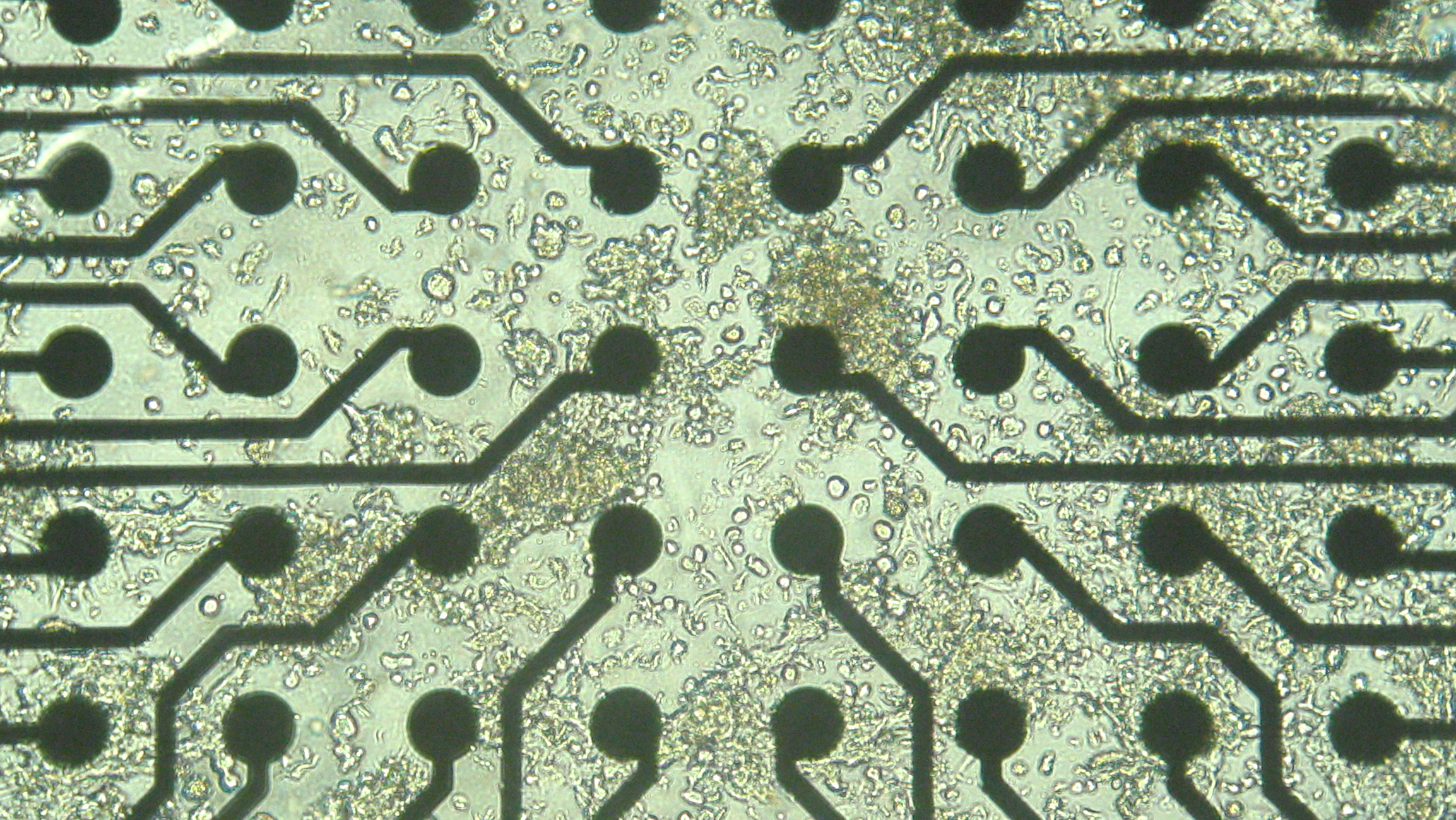}
\caption{IMR-32 cells in an MEA dish}
\label{fig:MEA}
\end{figure}

\begin{figure}
\centering
\includegraphics[width=1.5in]{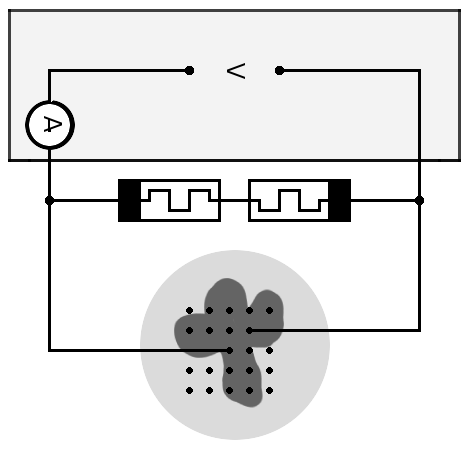}
\caption{Experimental scheme.}
\label{fig:Scheme}
\end{figure}

The memristors used in this study have two different behaviours~\cite{260}. Previously we reported that using a mixture of both types connected either in series with opposite polarity or in an anti-parallel configuration was most likely to lead to spiking dynamics~\cite{P0c}. For this study, a circuit of two memristors of different types in anti-polarity series was used, see figure~\ref{fig:Scheme}. The current was measured at 0.56s intervals for 560s using a Keithley sourcemeter (controlled via MatLab). that provides a voltage of +0.5V, for 560s, measuring current every 0.56s. 

The cells used are IMR-32 neuroblastoma cells~\cite{RB1}. The cells were cultured in RPMI medium (RPMI-1640, Sigma-R0883) with 10\% heat-inactivated fetal bovine serum (Sigma-F0804), 2mM L-glutamine (Sigma-G7513), essential amino acids (RPMI 1640 amino acids solution, Sigma-R7131) and 0.01\% penicillin-streptomycin (Sigma-P1333). The cells (~25,000 cells/dish) were transferred to a multi-electrode array (MEA) dish (Avenda Biosystems) coated with PEI~\cite{RB4} and incubated at physiological temperatures for between 24 and 48 hours. The MEA comprises 64 microelectrodes 30$\mu$M diameter and 200$\mu$M apart which allows the measurement of physiological voltages (typically $\le$ 200mV). The two electrodes in close proximity of each other with the largest amount of spiking activity were selected to be connected to the external memristor circuit.

Two different ways of promoting cell spiking were used: 1. A subset of the IMR-32 cells were placed in a medium with added [dibutyryl cAMP, 0.5mg/ml, Sigma-D0627], which causes differentiation and promotes spiking~\cite{RB2}; 2. 10$\mu$L of 130mM capsaicin (Sigma-M2028) in phosphate buffered saline was added 200s into experiments with either undifferentiated, or differentiated cells - capsaicin is known to promote TRPV1-dependent calcium signalling, and differentiated IMR-32 cells have been shown to express TRPV1~\cite{RB3}. The first dose of capsaicin was administered when the cells were not connected to the memristors to check that capsaicin caused increased spiking, then the memristors were connected for the second and third doses, further doses caused a negligible effect due to cell desensitization. 

Sixteen experiments will be discussed: 7 experiments with undifferentiated, unstimulated cells (with 3 control experiments comparing electrode separation) and 9 with stimulation of spiking by adding capsaicin, 3 of which were differentiated cells. For the memristor network control, the cables to the MEA dish were disconnected and left as an open-circuit. For the MEA dish controls, a clean MEA dish was filled with (in turn) distilled water, deionised water or RPMI medium as made up for the cells (and warmed to body temperature).

\section{Results}

\subsection{Controls}

\subsubsection{Isolated Memristor Network}

\begin{figure}
\centering
\subfigure[2-memristors without cells]{%
\includegraphics[width=1.5in]{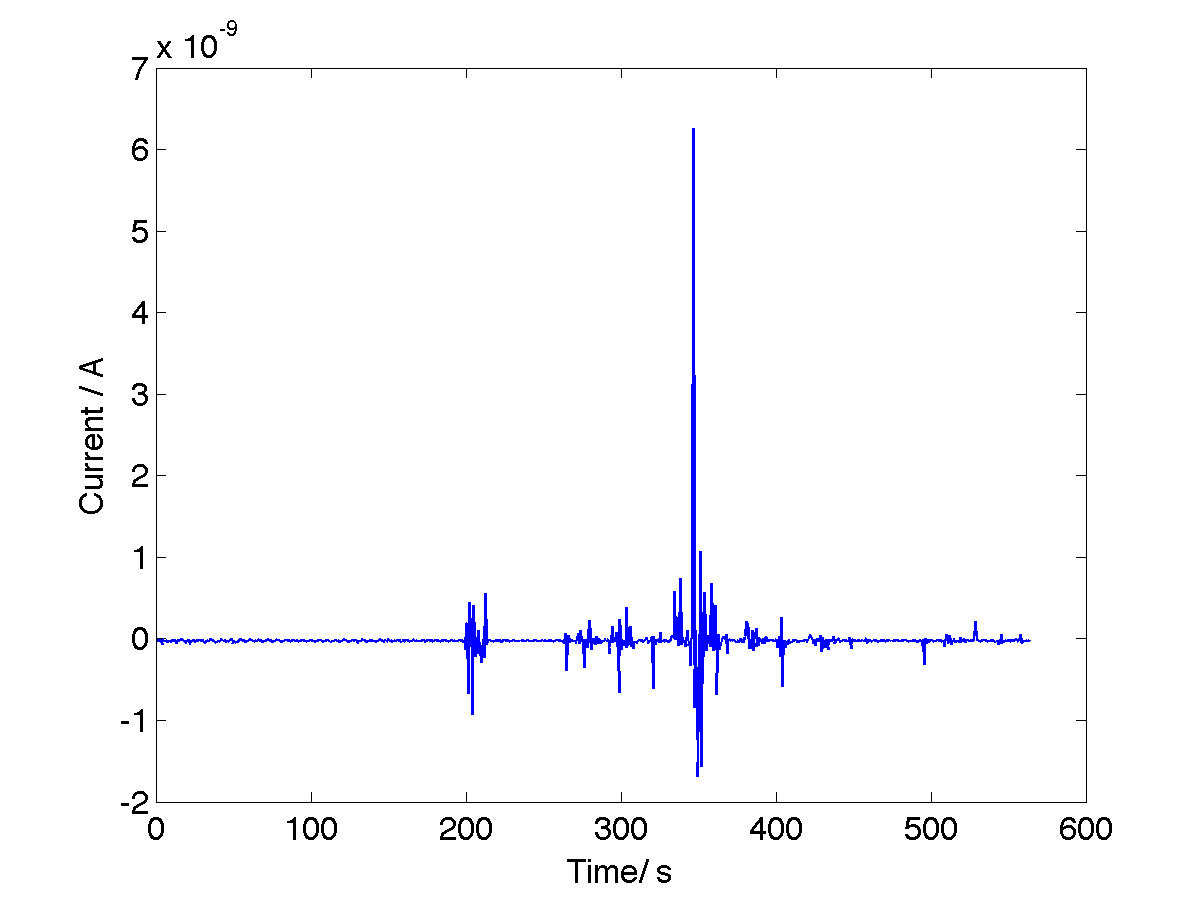}
\label{fig:Spiking}
}
\subfigure[cell medium only]{%
\includegraphics[width=1.5in]{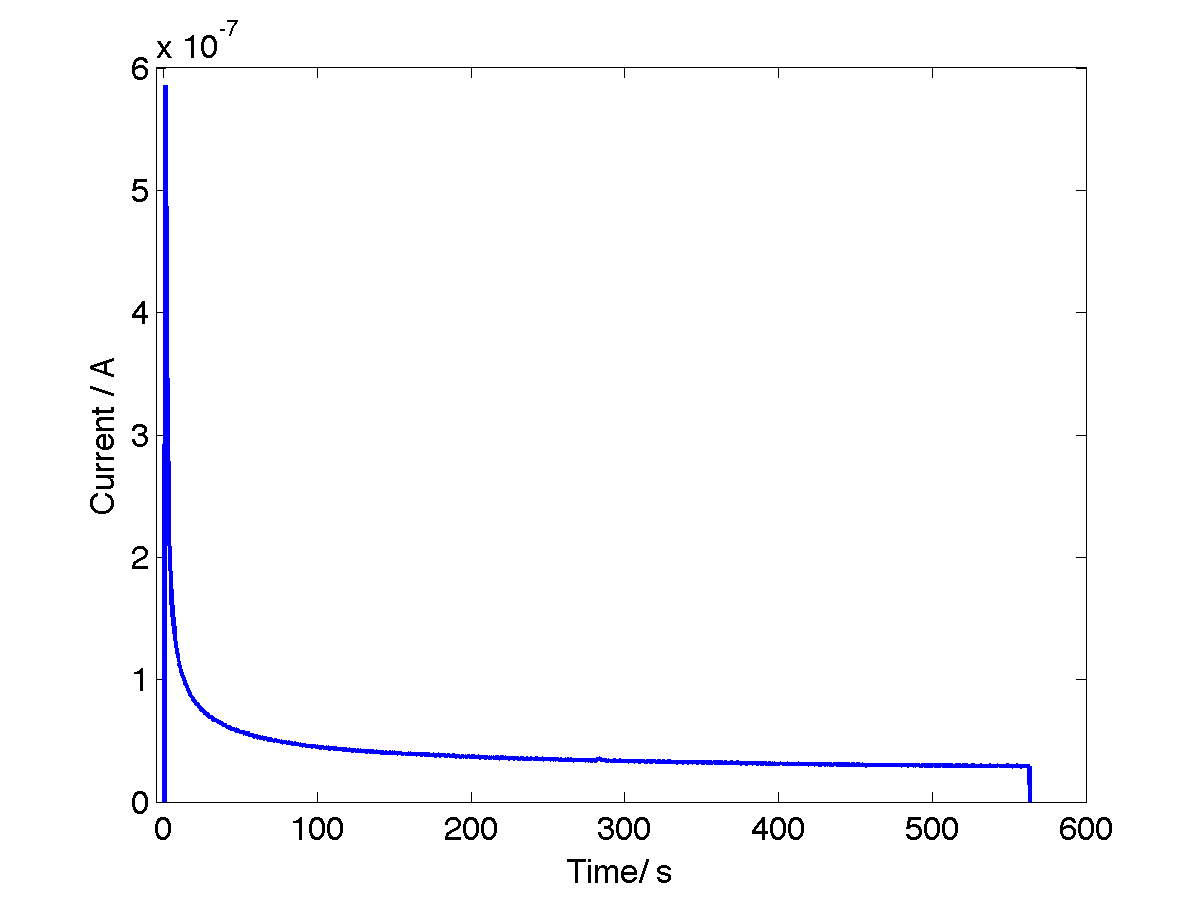}
\label{fig:dH20}
}
\caption{Control experiments}
\end{figure}

The responses of the memristor network when not connected to the cells is shown in figure~\ref{fig:Spiking}, this data was recorded after the cell spiking experiments however is qualitatively similar to that recorded beforehand (not shown) and in other spiking memristor networks~\cite{P0c}. This figure demonstrates expected 2-memristor network dynamics. A single memristor has a spike decay curve and a negative spike when the voltage is turned off, when combined into networks the starting (and ending) spikes are suppressed, instead the system shows sudden emergent bursting spikes, such as those seen after 200s in figure~\ref{fig:Spiking}.

\subsubsection{Memristor Network Connected to MEA Dish Without Cells}

Cell culture medium is an aqueous solution containing a mix of ions, sugars, colloids, fats and proteins, therefore any observed effect needs to be separated from that of connecting a memristor network to an aqueous sink of charged particles. Thus, we compared the results for an MEA dish filled with RPMI and distilled water. To completely remove the effects of ions on any conductivity, ultrapure deionised water was also tested. 

Results for the cell culture medium control are shown in~\ref{fig:dH20} and this response is the expected response for a single memristor~\cite{SpcJ} or a memristor network with low compositional complexity and `ideal' memristors. The other two controls (distilled water and deionised water) are identical and all start with the current of 60$\mu$A (a measured resistance of 8333$\Omega$). The memristor spiking is suppressed and it appears that the current spikes are removed from the memristor network or otherwise damped by connection to a large sink of ions. The control experiments show that this effect is due to the loss of current from the networks to the aqueous medium, and not to cellular activity. Thus, it seems that connection to an external current sink removes the non-idealities of a memristor network (not permanently, because the network still exhibits standard memristor network dynamics after the experiments). We believe that the memristor network spiking is due to spike-response-caused voltage 
changes across individual memristors being out of sync (as postulated in~\cite{Mu0}). If this is hypothesis is correct, then the connection to a sink of current would likely prevent this mechanism occurring by damping the system dynamics.

\subsection{The Effect of Coupling to an External Pool of Spiking Neurons}

\begin{figure}
\centering
\includegraphics[width=3.3in]{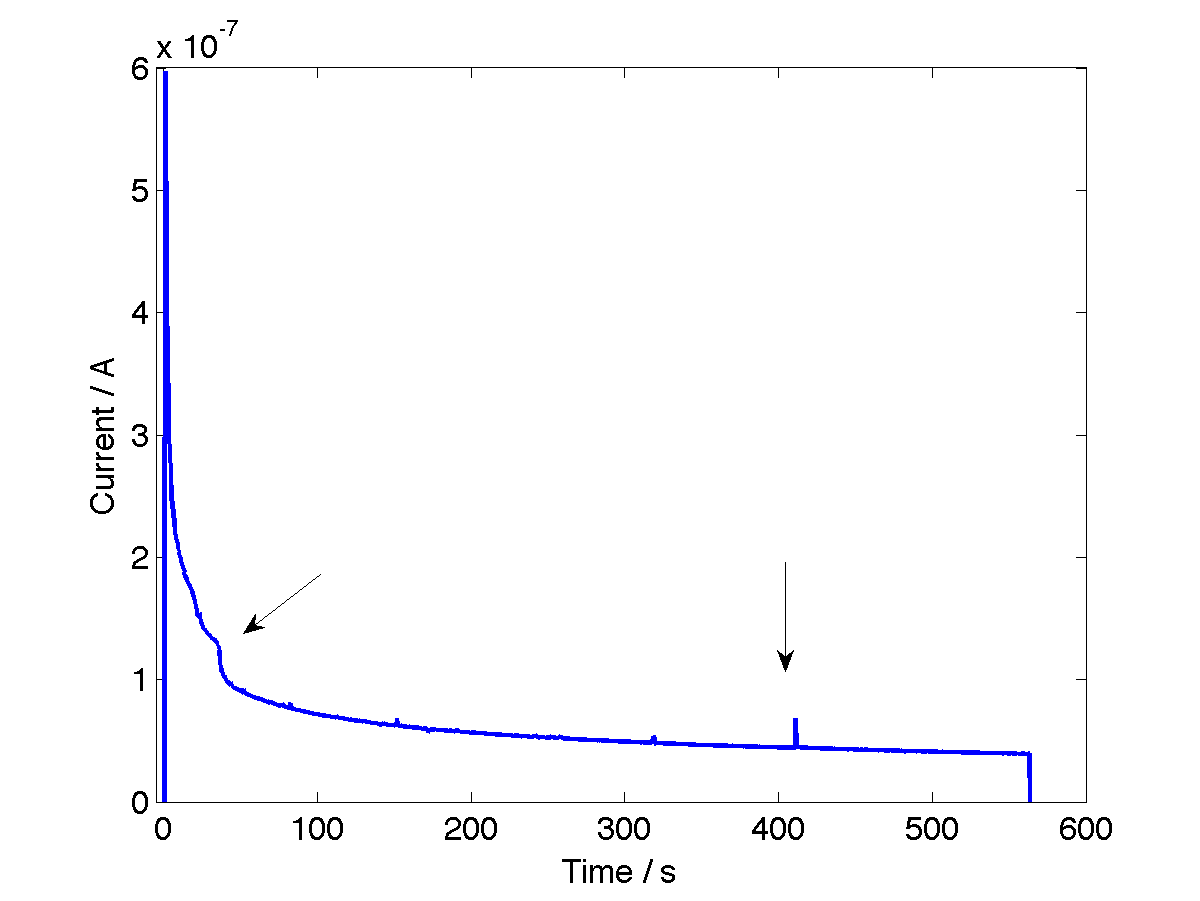}
\caption{The effect of cells. Measured $I-t$ curve for the 2-memristor circuit. Arrows indicate the possible effect of cellular processes}
\label{fig:CT1}
\end{figure}

Figure~\ref{fig:CT1} shows the effect on the memristor $I-t$ profile of coupling the 2-memristor circuit to an MEA dish full of living, spiking neuronal cells. The memristor spiking is again suppressed, as with the control experiment involving medium. However, by comparing figure~\ref{fig:CT1} with the control in figure~\ref{fig:dH20}, we can see that there is a difference due to the action of the cells: there is a discontinuity at ~30s and a spike seen at ~410s, as marked by arrows in figure~\ref{fig:CT1}. The discontinuity is similar to that seen when a voltage changes across the memristor and could be the result of a voltage spike changing the `state' of the memristor, the spike may be caused by the cells themselves or a memristor spike retained in the network.

\begin{figure}
\centering
\includegraphics[width=3.3in]{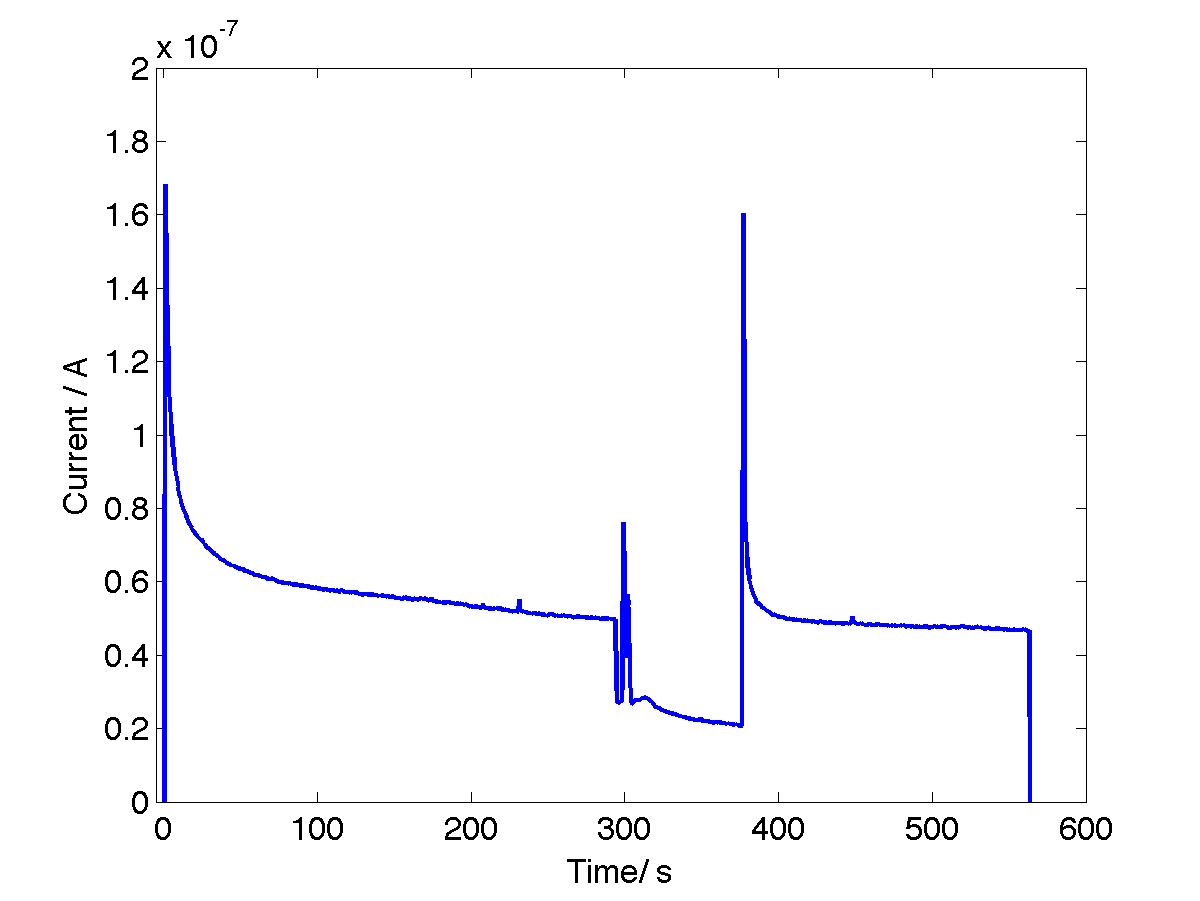}
\caption{Repeat of experiment, using a dish containing more cells and after an extra day's growth.}
\label{fig:CT2}
\end{figure}

Figure~\ref{fig:CT2} shows the same experiment for a different dish that contained more cells, had been incubated for an extra day and  where the electrodes were closer (neighbours along the edges of the rectilinear grid rather than on the diagonal). This data clearly shows an effect of coupling to the cells at ~300 and 400s. The spike at 400s is consistent to a memristor being switched to a higher voltage, and is suggestive of a cellular change altering the resistance of the memristor. The drop and spike at around 300s is not consistent with this, and could be the memrsitor network reacting to neuronal dynamics. 

A full set of control experiments for electrode separation (not shown) have been run comprising of 1,2 and 3 diagonals, equivalent to 283$\mu$m, 566$\mu$m and 849$\mu$m, which demonstrate that the amount of perturbation of the memristor's $I-t$ curve is increased for shorter distances, i.e. the closer the electrodes, the shorter distance any signal has to travel to affect the exterior memristor circuit, thus the more likely it is that cellular processes will affect the memristors. 

\subsection{The Effect of Extra Spiking}

\begin{figure}
\centering
\includegraphics[width=3.3in]{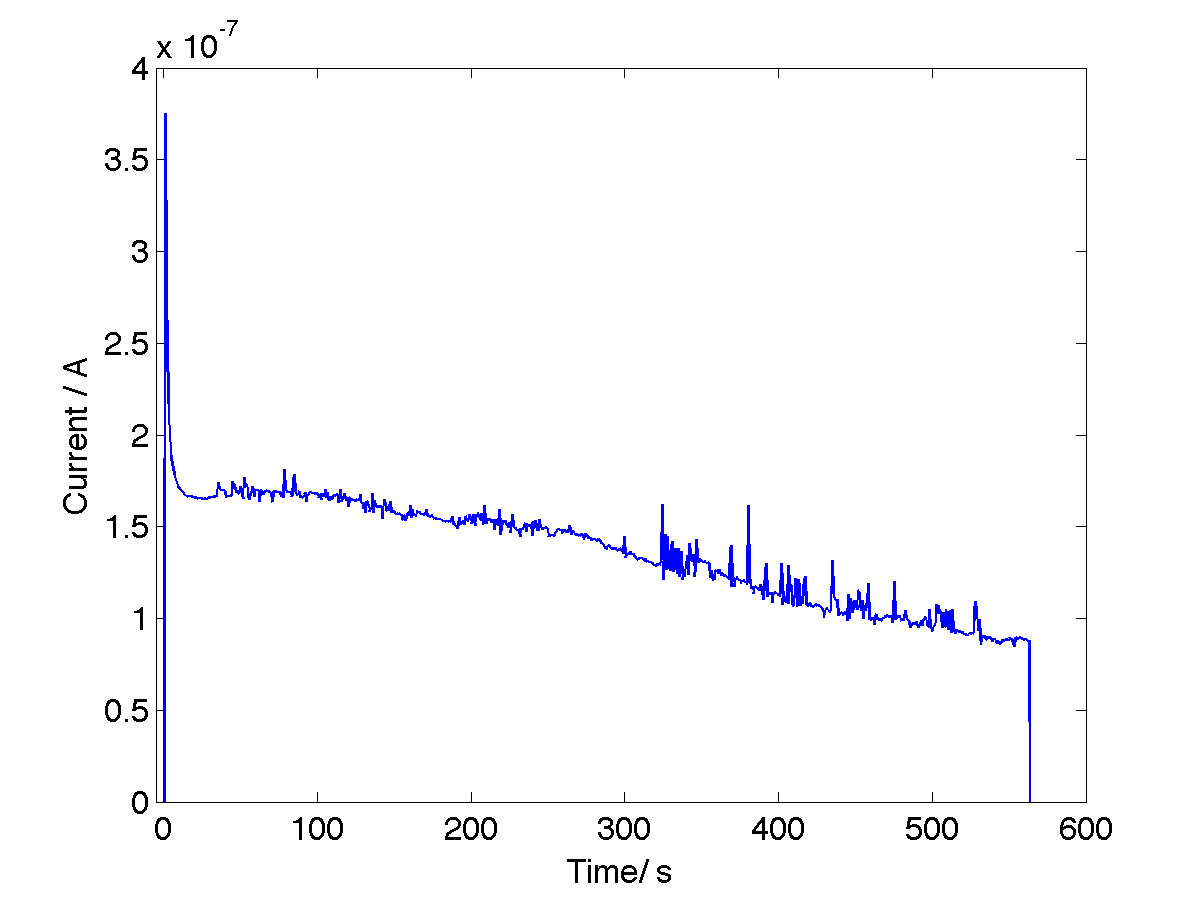}
\caption{Capsaicin added whilst memristors were connected.}
\label{fig:CapTest}
\end{figure}

Figure~\ref{fig:CapTest} show the effects of adding capsaicin to chemically stimulate cell spiking. The cells were subject to 10$\mu$L at approximately 10 minutes before starting the experiment and a second amount at 200s into this run. The main result is that increased cell activity has greatly changed the shape to a sloping background line with many spikes superimposed on it. This shows that an external memristor circuit can `record' spiking activity from an electrically connected living neuronal network. The second result is that the effect of the increased capsaicin dose can be seen ~110s after it was added in the form of an increased number and magnitude of the spikes.

\subsection{Behaviours observed}

\begin{table}
\renewcommand{\arraystretch}{1.3}
\caption{The number of each type of graph from the preliminary results. *Differentiated results all had added capsaicin.}
\label{tab:Results}
\begin{center}
\begin{tabular}{|c||c|c|c|c|c|c|}
\hline
Behaviour 	& Curve	& Curve	& Curve	& Curve 	& Line & Line\\
				& no 		& few 		& with 	&with 		& few 	& many	\\
				& spikes	& spikes 	& steps 	&switch	& spikes & spikes	\\
\hline
Total no. 			& 2 		& 2 		& 4	 		& 1  		& 1 	& 6\\
Undifferentiated		& 2 		& 0 		& 1		   	& 1			& 1		& 2\\
With capsaicin& 0		& 2			& - 		& 0			& 0 	& 4\\
Differentiated* & 0		& 0 		& 3			& 0 		& 0		& 0\\
\hline
\end{tabular}
\end{center}
\end{table}

The types of behaviour were classified into 6 behaviours based on visual appearence (a full analysis of these will appear in a forthcoming paper). The expected single memristor reponse, as shown in figure~\ref{fig:dH20} is labelled as a `curve no spikes' in table~\ref{tab:Results}, the same curve with spikes superimposed (not shown), is labelled `curve few spikes', a curve with a change in gradient like the step shown in figure~\ref{fig:CT1} is `curve with steps' (multiple steps were seen in some cases), figure~\ref{fig:CT2} which seems to show a change in voltages applied to a memristor is labelled `curve with switch'. Anything that was not approximately similar to the d.c. spike shape observed in figure~\ref{fig:dH20}, was labelled a `line' and there were several roughly sloping lines, these are split by a estimation of the amount of spiking, with figure~\ref{fig:CapTest} an example of a line with many spikes.

There are 16 individual data sets compared, which is enough to give an indication of an effect, but too few for statistical analysis. Bearing this in mind, we can see that the step superimposed with a curve is likely in differentiatd cells (with a capsaicin-induced increase in spiking rate); conversely, the addition of capsaicin to undifferentiated cells is more likely to lead to a line with spikes. A standard curve with spikes overlaid was only seen under capsaicin, implying that these spikes come from the cells and not the memristors themselves (a capsaicin and water control was run and gave the same results for water). It is possible for some experiments to exhibit no spikes in the memristor circuit at all, even though the memsitors were only connected to areas that were seen to be actively spiking, so in those cases cell spiking activity either stopped or was not transmitted to the memristor circuit. Finally, the cells with no chemical additives to induce spiking seem to have a range of behaviours as 
measured by the memristor sub-circuit.

\section{Conclusions}

This work demonstrates that memristors can be switched by neuronal cell action, which is exciting as it could allow the brain to communicate with external electronics via a memristor circuit. Furthermore, we've shown that connecting spiking neuronal cells to a spiking memristor changes the dynamics in a qualifiable way, and related some of these changes to extra spiking activity.

The cells tested are human neuroblastoma cells, and although they exhibit spiking properties, they do not behave exactly like normal neurons. We are currently repeating this work with ND7/23 cell-line which is more thoroughly characterised in neuronal studies and isn't derrived from cancer. The supporting data from the MEA dishes requires more runs in order to get statistical samples (to answer the question of whether the memristor spiking causes an increase in cell firing rate), so these experiments are being repeated.

Finally, it is expected that a 2-D network will have different properties to a 3-D network, and that a large part of the brain's properties and function require the extra connectivity which requires an extra dimension, therefore we are currently generating 3-D spheroids from ND7/23 cells. 

\bibliographystyle{IEEEtran.bst}

\maketitle


\end{document}